\def\be{\begin{equation}}
\def\ee{\end{equation}}
\def\ba{\begin{array}}
\def\ea{\end{array}}

\documentclass[prl,showpacs,twocolumn,amsmath]{revtex4}
\usepackage{mathrsfs}
\usepackage{amssymb}
\usepackage{pifont}
\usepackage{epsfig,subfigure,dsfont,amsthm,amsbsy,mathrsfs,amscd}
\def\qed{\leavevmode\unskip\penalty9999 \hbox{}\nobreak\hfill
     \quad\hbox{\leavevmode  \hbox to.77778em{%
               \hfil\vrule   \vbox to.675em%
               {\hrule width.6em\vfil\hrule}\vrule\hfil}}
     \par\vskip3pt}

\newtheorem{theorem}{Theorem}

\input amssym.def
\begin{document}
\title{\large\bf Sharing quantum nonlocality and genuine nonlocality
with independent observables}

\author{ Tinggui Zhang$^{1, \dag}$ and Shao-Ming Fei$^{2,3,\sharp}$}
\affiliation{ ${1}$ School of Mathematics and Statistics, Hainan Normal University, Haikou, 571158, China \\
$2$ School of Mathematical Sciences, Capital Normal University, Beijing 100048, China \\
$3$ Shenzhen Institute for Quantum Science and Engineering, Southern University of Science and Technology, Shenzhen 518055, China\\
$^{\dag}$ Correspondence to tinggui333@163.com\\
$^{\sharp}$ Correspondence to feishm@cnu.edu.cn}
\bigskip

\begin{abstract}

Recently the authors in [Phys. Rev. Lett. 125, 090401 (2020)]
considered the following scenario: Alice and Bob each have half of a
pair of entangled qubit state. Bob measures his half and then passes
his part to a second Bob who measures again and so on. The goal is
to maximize the number of Bobs that can have an expected violation
of the Clauser-Horne-Shimony-Holt (CHSH) inequality with the single
Alice. By taking the maximally entangled pure two-qubit state
$|\phi\rangle=\frac{1}{\sqrt{2}}(|00\rangle+|11\rangle)$ as an
example, it has been constructively proved that arbitrarily many
independent Bobs can share the nonlocality with the single Alice.
Here we demonstrate that arbitrarily many independent observers can
share the nonlocality of a single arbitrary dimensional bipartite
entangled but not necessary two-qubit entangled state. Further,
taking the generalized GHZ states as an example, we show that at
most two Charlies can share the genuine nonlocality of a single
generalized GHZ state with an Alice and a Bob. \end{abstract}

 \maketitle

\section{Introduction}
Quantum nonlocality, such as that revealed by violating the Bell
inequalities of quantum entangled states \cite{jsbe}, is one of the
most startling predictions of quantum mechanics. Recently, as
confirmed in loophole-free experiments \cite{bmlk}, nonlocality has
been proven to be useful in many quantum tasks such as
device-independent cryptography \cite{aanb} and randomness
certification \cite{spaa,rcol,lkyz,wlml}. A successful and secure
quantum network relies on quantum correlations distributed and
shared among many sites \cite{hjki}. Different kinds of multipartite
quantum correlations have been considered as valuable resources for
various applications in quantum communication tasks. A key property
is that such quantum correlations cannot be freely shared among the
multipartite systems, see e.g. \cite{monogamy} and references therein.

Recently, in Ref \cite{rsng,smam,assd,dass,pjbr}, the authors
studied the fundamental limits on nonlocality, asking whether a
single pair of entangled qubits could generate a long sequence of
nonlocal correlations. This sequential scenario (see FIG. 1) was
introduced first in \cite{rsng}. With the same sharpness of the two
measurements applied by each Bob, in Ref. \cite{smam} the authors
shown that at most two Bobs can achieve an expected CHSH \cite{chsh}
violation with a single Alice, in line with the numerical evidence
from \cite{rsng}. Equal sharpness two-outcome measurements were also
adopted in \cite{assd} to show that at most two Bobs can share the
Bell nonlocality of a maximally entangled state with a single Alice
in this scenario. It has been shown that
at most two Bobs can exhibit bipartite nonlocality with a single
Alice by using local realist inequalities with three and four
dichotomic measurements pre observer \cite{dass}. More recently, in \cite{pjbr} the authors
studied such scenario and shown that if the Bobs' apply different
measurements, then arbitrarily many independent Bobs can share the
nonlocality of the maximally entangled pure two-qubit state
$|\phi\rangle=\frac{1}{\sqrt{2}}(|00\rangle+|11\rangle)$ with the
single Alice.
\begin{figure}[ptb]
\includegraphics[width=0.45\textwidth]{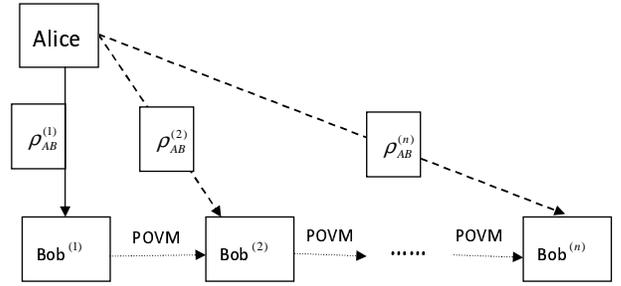}\caption{A quantum state $\rho_{AB}^{(1)}$ is initially shared  by Alice and Bob$^{(1)}$. Bob$^{(1)}$ performs first a measurement
on his part and then passes it to Bob$^{(2)}$. The post-measurement
state is $\rho_{AB}^{(2)}$. Bob$^{(2)}$ measures $\rho_{AB}^{(2)}$
on his part and passes it to Bob$^{(3)}$ and so on.}
\end{figure}

High-dimensional quantum entanglement and nonlocality provide a playground for fundamental research and also lead to technological advances, with stronger locality violations that can be exploited to tolerate larger amounts of noise in quantum communication protocols.
Various physical and technical approaches on how to manipulate multilevel quantum states in different degrees of freedom have been presented, inspiring new synergies that create new technologies such as teleporting the complete quantum information stored in a single ``photon" \cite{zeilinger}. Therefore, in this article we study such nonlocal
correlation sharing scenario for arbitrary high dimensional
bipartite entangled pure states. We show that arbitrarily many
independent observers can share the nonlocality of any single
arbitrary dimensional bipartite entangled states. Furthermore, we
investigate the genuine nonlocality sharing among Alice, Bob and
Charlies, see FIG. 2. By using the Svetlichny inequality
\cite{svet}, we show that at least two Charlies can share the
genuine nonlocality of a single generalized GHZ state with Alice and
Bob.

\begin{figure}[ptb]
\includegraphics[width=0.45\textwidth]{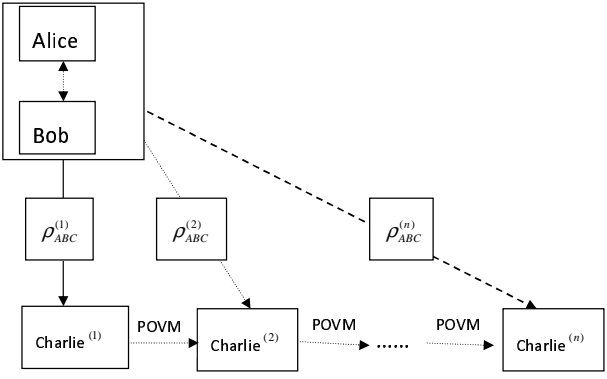}\caption{
A quantum state $\rho_{ABC}^{(1)}$ is initially shared by Alice, Bob
and Charlie$^{(1)}$. Charlie$^{(1)}$ performs a measurement on her
part and passes it to Charlie$^{(2)}$. The post-measurement state is
$\rho_{ABC}^{(2)}$. Charlie$^{(2)}$ measures $\rho_{ABC}^{(2)}$ on
her part and passes it to Charlie$^{(3)}$ and so on.}
\end{figure}

\section{Nonlocal sharing of bipartite high-dimensional pure states}

We first consider the
measurement scenario such that Alice attempts to share the nonlocal correlations of an entangled pure state with $n$ independent Bobs (FIG. 1).
Any bipartite pure state $|\psi\rangle\in H_A\otimes H_B$ with
$dim(H_A)=s$ and $dim(H_B)=t$ ($s \leq t$) has Schmidt
decomposition form, $|\psi\rangle=\sum_{i=1}^sc_i|i_A\rangle|i_B\rangle$,
where $c_i\in [0,1]$, $\sum_i^s c_i^2=1$, $\{i_A\}_1^s$ and
$\{i_B\}_1^t$ are the orthonormal bases of $H_A$ and $H_B$, respectively.
$|\psi\rangle$ is entangled if and only if at least two $c_i$s are
nonzero. Without loss of generality, below we assume that
$c_i$ are arranged in descending order.

To begin with, Alice and Bob$^{(1)}$ share an arbitrary entangled
bipartite pure state
$\rho_{AB}^{(1)}=|\psi\rangle\langle\psi|=\Sigma_{i,j}c_ic_j|ii\rangle\langle
jj|$. Bob$^{(1)}$ proceeds by choosing a uniformly random input,
performing the corresponding measurement and recording the outcome.
Denote the binary input and output of Alice (Bob$^{(k)}$) by $X$
($Y^{(k)}$) and $A$ ($B^{(k)}$), respectively. Suppose Bob$^{(1)}$
performs the measurement according to $Y^{(1)}=y$ with the outcome
$B^{(1)}=b$. Averaged over the inputs and outputs of Bob$^{(1)}$,
the unnormalized state shared between Alice and Bob$^{(2)}$ is given
by $$
\rho_{AB}^{(2)}=\frac{1}{2}\Sigma_{b,y}(I_s\otimes\sqrt{B_{b|y}^{(1)}})
\rho_{AB}^{(1)}(I_s\otimes\sqrt{B_{b|y}^{(1)}}) $$ where
$B^{(1)}_{b|y}$ is the positive operator-valued measure (POVM)
effect corresponding to outcome $b$ of Bob$^{(1)}$'s measurement for
input $y$, $I_s$ is the $s\times s$ identity matrix. Repeating this
process, one gets the state $\rho_{AB}^{(k)}$ shared between Alice
and Bob$^{(k)}$.

To detect the nonlocality we employ the CHSH inequality \cite{chsh},
$I_{CHSH}=\langle \mathbb{B}\rangle \leq 2$, where $\langle \mathbb{B} \rangle =
Tr(\mathbb{B}\rho)$, $\mathbb{B}=A_0\otimes B_0+A_0\otimes B_1+A_1\otimes
B_0-A_1\otimes B_1$, $A_i$ and $B_i$, $i=0,1$, are Hermitian operators with
eigenvalues $\in[-1,1]$. If for some binary observables $A_i$ and
$B_i^{(k)}$, $i=0,1$, $I_{CHSH}^{(k)}\equiv Tr(\mathbb{B}\rho_{AB}^{(k)})>2$,
then the state $\rho_{AB}^{(k)}$ is nonlocally correlated.

For the case that $s$ and $t$ are even, we employ the
POVMs with measurement operators $\{E,I-E\}$, where $E$ has the form
$E=\frac{1}{2}(I_m+\gamma(I_{\frac{m}{2}}\otimes{\vec{r}}\cdot\vec{\sigma}))$,
$\vec{r}\in R^3$ with $\|\vec{r}\|=1$,
$\vec{r}\cdot\vec{\sigma}=r_1\sigma_1+r_2\sigma_2+r_3\sigma_3$, $\sigma_i$, $i=1,2,3$, are the
standard Pauli matrices, $\gamma \in [0,1]$ is the sharpness of the measurement, $I_m$ stands for the $m\times m$ identity matrix, $m=s,t$.
We set the Alice's POVMs to be

\be\label{a00o} A_{0|0}=\frac{1}{2}(I_s+(I_{\frac{s}{2}}
\otimes(\cos\theta\sigma_3+\sin\theta\sigma_1)), \ee \be\label{a01o}
A_{0|1}=\frac{1}{2}(I_s+(I_{\frac{s}{2}}
\otimes(\cos\theta\sigma_3-\sin\theta\sigma_1)), \ee for some
$\theta\in (0,\frac{\pi}{4}]$. For each $k=1,2,\cdots,n$,
Bob$^{(k)}$'s POVMs are defined as \be\label{b00o}
B^{(k)}_{0|0}=\frac{1}{2}(I_t+(I_{\frac{t}{2}}\otimes(\cos\theta\sigma_3)),
\ee \be\label{b01o}
B^{(k)}_{0|1}=\frac{1}{2}(I_t+(I_{\frac{t}{2}}\otimes(\gamma_k\sin\theta\sigma_1)),
\ee $k=1,2,\cdots,n$.

When $s$ and $t$ are odd, we employ the
POVM measurement operators $\{E,I-E\}$, with
$$E=\frac{1}{2}\left[I_m+\gamma\left(\begin{array}{cc}
    I_{[\frac{m}{2}]}\otimes{\vec{r}}\cdot\vec{\sigma} &  0 \\
    0 & 1
\end{array}\right)\right],
$$
where $m=s,t$, $[p]$ represents the integer less or equal to $p$.
The Alice's POVMs are defined as \be\label{a00e}
A_{0|0}=\frac{1}{2}\left[I_s+\left(\begin{array}{cc}
    I_{[\frac{s}{2}]}\otimes(\cos\theta\sigma_3+\sin\theta\sigma_1) &  0 \\
    0 & 1 \\
  \end{array}\right)\right],
\ee \be \label{a01e}
A_{0|1}=\frac{1}{2}\left[I_s+\left(\begin{array}{cc}
    I_{[\frac{s}{2}]}\otimes(\cos\theta\sigma_3-\sin\theta\sigma_1) &  0 \\
    0 & 1 \\
  \end{array}\right)\right]
\ee for some $\theta\in (0,\frac{\pi}{4}]$. The Bob$^{(k)}$'s POVMs
set to be \be\label{b00e}
B^{(k)}_{0|0}=\frac{1}{2}\left[I_t+\left(\begin{array}{cc}
    I_{[\frac{t}{2}]}\otimes\sigma_3 &  0 \\
    0 & 1 \\
  \end{array}\right)\right],
\ee \be\label{b01e}
B^{(k)}_{0|1}=\frac{1}{2}\left[I_t+\gamma_k\left(\begin{array}{cc}
    I_{[\frac{t}{2}]}\otimes\sigma_1 &  0 \\
    0 & 1 \\
\end{array}\right)\right],
\ee $k=1,2,\cdots,n$.

The observables are then given by $A_x=A_{0|x}-A_{1|x}$ and
$B_y^{(k)}=B_{0|y}-B_{1|y}$, $x,y=0,1$, for both even and odd $t,s$.
We have the following conclusion for the expected CHSH value for Alice and Bob$^{(k)}$,
see proof in Appendix.

\begin{theorem} For any initial entangled bipartite pure quantum state
$|\psi\rangle\in H_A\otimes H_B$ with Schmidt decomposition
$|\psi\rangle=\sum^s_{i=1}c_i|i_A\rangle|i_B\rangle$, the expected
CHSH value of $\rho_{AB}^{(k)}$ is given by \be\label{s3}
I^{(k)}_{CHSH}\geq 2^{2-k}\left[\gamma_k
L\sin\theta+\cos\theta\prod_{j=1}^{k-1}(1+\sqrt{1-\gamma_j^2})\right],
\ee where
$L=2(c_1c_2+c_3c_4+\cdots+c_{2[\frac{s}{2}]-1}c_{2[\frac{s}{2}]})$.
\end{theorem}

Next we show that there exist suitable parameters $\gamma_k$ and
$\theta$ such that $I_{CHSH}^{(k)}>2$ for arbitrary $k$.
From Theorem 1 we require that
\begin{eqnarray}\label{s10}& &\gamma_k>\frac{2^{k-1}-\cos\theta\prod_{j=1}^{k-1}(1+\sqrt{1-\gamma_{j}^2})}{L\sin\theta}.
\end{eqnarray}
Set $\gamma_1(\theta)=(1+\epsilon)\frac{1-\cos(\theta)}{L\sin(\theta)}$ for $\varepsilon > 0$.
One has
\begin{eqnarray}\label{s11}
& &\gamma_k(\theta)=(1+\epsilon)\frac{2^{k-1}-\cos\theta\prod_{j=1}^{k-1}
(1+\sqrt{1-\gamma_{j}^2})}{L\sin\theta}
\end{eqnarray}
for $\gamma_{k-1}(\theta)\in(0,1)$, $k\in \{2,\cdots,n\}$.

\begin{theorem} For each $n\in \mathbb{N}$, there exists a sequence
$\{\gamma_k\}_1^n$ and $\theta_n\in(0,\frac{\pi}{4}]$ such that
$I_{CHSH}^{(k)}>2$ for all $k=1,2,\cdots,n$, with $\theta\in(0,\theta_n)$
and $\gamma_k(\theta) < 1$ for all $k\leq n$.
\end{theorem}

The proof is given in Appendix. Theorem 2 shows that arbitrarily
many independent observers can share the nonlocality of a single
arbitrary dimensional bipartite entangled state, as long as at least
two Schmidt coefficients $c_i$ of the state are nonzero. Moreover,
the state is not necessary to be maximally entangled.

\section{Genuine nonlocal sharing of three-qubit states}

Multipartite nonlocal correlations have not only foundational
implications \cite{ngis} but also novel applications in quantum
communication and quantum computation \cite{msju,msee,pzet,cylu}, as
well as in phase transitions and criticality in many-body systems
\cite{pzet}. Essentially different from the bipartite case, one has
so called genuine multipartite nonlocality for multipartite systems.
In the following, we consider the sharing ability of the genuine
nonlocality of the three-qubit generalized GHZ state, $
|\psi_{\alpha}\rangle=\cos\alpha|000\rangle+\sin\alpha|111\rangle. $

Let $A_i=\vec{a}_i\cdot\vec{\sigma}$, $B_i=\vec{b}_i\cdot\vec{\sigma}$ and $C_i=\vec{c}_i\cdot\vec{\sigma}$, $i=0,1$, be the measurement observables on the first, second and third qubit, respectively,
 with $\vec{a}_i$, $\vec{b}_i$ and $\vec{c}_i$ the real unit vectors. The Svetlichny operator is defined by \cite{svet},
\begin{eqnarray}\label{s12}& &S=A_0(B_0+B_1)C_0+A_0(B_0-B_1)C_1\nonumber\\& &+A_1(B_0-B_1)C_0-A_1(B_0+B_1)C_1.
\end{eqnarray}
If a state $|\psi\rangle$ admits bi-local hidden variable model,
then the expectation value of the Svetlichny operator satisfies the
Svetlichny's inequality,
\begin{eqnarray}\label{sineq}
\mathbb{S}(\psi)=\langle\psi|S|\psi\rangle\leq 4.
\end{eqnarray}
If (\ref{sineq}) is violated, $|\psi\rangle$ must be a genuine
three-qubit nonlocally correlated state. The maximal violation of
the Svetlichny inequality (\ref{sineq}) for the state
$|\psi_g\rangle$ has been studied in \cite{sgns}. It has been shown
that when $\sin^22\alpha>\frac{1}{2}$, the state $|\psi_g\rangle$ is
genuine three-qubit nonlocal.

We consider the following measurement scenario, see FIG. 2: Alice
and Bob want to share the genuine three-qubit nonlocality of single
three-qubit state $|\psi_g\rangle$ with possible $n$ independent
Charlies. Denote the binary input and output of Alice (Bob) by $X$
($Y$) and $A$ ($B$), respectively. For each $k\in {N}$ we denote the
binary input and output of Charlie$^{(k)}$ by $Z^{(k)}$ and
$C^{(k)}$, respectively. At the beginning, the three-qubit state
$|\psi_g\rangle$ is shared among Alice, Bob and Charlie$^{(1)}$.
Charlie$^{(1)}$ proceeds by choosing a uniformly random input,
performing the corresponding measurement and recording the outcome.
The postmeasurement qubit is then sent to Charlie$^{(2)}$. Suppose
Charlie$^{(1)}$ performs the measurement according to $Z^{(1)}=z$
and receives the outcome $C^{(1)}=c$. The postmeasurement state can
be described by the L\"uders rule. Averaged over the inputs and
outputs of Charlie$^{(1)}$'s, the postmeasurement unnormalized state
$\rho_{ABC}^{(2)}$ shared among Alice, Bob and Charlie$^{(2)}$ is
given by
\begin{eqnarray}
\rho_{ABC}^{(2)}=\frac{1}{2}\sum_{c,z}(I_2\otimes I_2\otimes\sqrt{C_{c|z}^{(1)}})\rho_{ABC}^{(1)}(I_2\otimes I_2\otimes\sqrt{C_{c|z}^{(1)}}),\nonumber
\end{eqnarray}
where $C^{(1)}_{c|z}$ is the POVM effect corresponding to the
outcome $c$ of Charlie$^{(1)}$'s measurement for input $z$.
Repeating this process, one can compute the state $\rho_{ABC}^{(k)}$
shared among Alice, Bob and Charlie$^{(k)}$. The expected value of
Svetlichny operator associated with the state $\rho_{ABC}^{(k)}$ and
the binary random observables $A_i$, $B_i$ and $C_i^{(k)}$, $i=0,1$,
is given by $S^{(k)}\equiv Tr(S\rho_{ABC}^{(k)})$.

To see the maximal $k$ such that $S^{(k)} > 4$, we consider again two-outcome POVMs
$\{E,I-E\}$. Let Alice's POVMs be given by
$$
A_{0|0}=\frac{1}{2}(I+\sigma_1), ~~~
A_{0|1}=\frac{1}{2}(I+\sigma_2)
$$
and Bob's POVMs by
$$
\begin{array}{rcl}
B_{0|0}&=&\displaystyle\frac{1}{2}(I+\cos\theta\sigma_1-\sin\theta\sigma_2),\\[2mm]
B_{0|1}&=&\displaystyle\frac{1}{2}(I+\cos\theta\sigma_1+\sin\theta\sigma_2)
\end{array}
$$
for some $\theta\in (0,\frac{\pi}{4}]$.
The Charlie$^{(k)}$'s POVMs are defined by
$$C^{(k)}_{0|0}=\frac{1}{2}(I+\sigma_1),~~~
C^{(k)}_{0|1}=\frac{1}{2}(I+\gamma_k\sigma_2)
$$
for $k=1,2,\cdots,n$.

Set $A_x=A_{0|x}-A_{1|x}$, $B_y=B_{0|y}-B_{1|y}$ and $C^{(k)}_z=C^{(k)}_{0|z}-C^{(k)}_{1|z}$, $x,y,z=0,1$. We have the following conclusion, see detailed proof in Appendix.

\begin{theorem}
For the initially shared  generalized GHZ state
$|\psi_{\alpha}\rangle$, the expected value of the Svetlichny
operator with respect to the state $\rho_{ABC}^{(k)}$ is given by
\be\label{s13}
S^{(k)}=2^{2-k}\sin2\alpha(\cos\theta+\sin\theta)(\gamma_k
+\prod_{j=1}^{k-1}(1+\sqrt{1-\gamma_j^2})). \ee
\end{theorem}

For $k=1$, we have $S^{(1)}=2\sin2\alpha(\cos\theta+\sin\theta)(\gamma_k +1)$. If
$\gamma_k=1$ and $\theta=\frac{\pi}{4}$, then $S^{(1)}>4$ as long as
$\sin^2(2\alpha) >\frac{1}{2}$. By detailed analysis, we have, see proof in Appendix,

\begin{theorem}
There are at most two Charlies sharing the genuine nonlocality with
Alice and Bob, $S^{(k)}>4$, $k=1,2$, when $\sin^22\alpha\in
(\frac{8}{9},1]$ for the initially shared  generalized GHZ state
$|\psi_{\alpha}\rangle$.
\end{theorem}

In \cite{sdsd} the authors considered this problem from the
formalism of weak or unsharp measurements for the GHZ stat,
$|GHZ\rangle=\frac{1}{\sqrt{2}}(|000\rangle+|111\rangle)$. Our
conclusion is for the generalized GHZ states: when $\sin^22\alpha\in
(\frac{8}{9},1]$, at most two Charlies can share the three-qubit
genuine nonlocality of a single generalized GHZ state with an Alice
and a Bob, which coincides with the result derived in [24] as a
special case ($\alpha=\frac{\pi}{4}$).

\section{Conclusions and discussions}
Quantum nonlocality is a fundamental feature in quantum mechanics.
We have demonstrated that it is possible for arbitrarily many
independent Bobs to violate the CHSH inequality with a single Alice
by using any bipartite high-dimensional pure states that are either
maximally or non-maximally entangled. As the approach used for qubit
case can not be used for high dimensional case since relations like
$Tr(\rho(\sigma_{\vec{a}}\otimes\sigma_{\vec{b}}))=(\vec{a},T(\rho)\vec{b})$
\cite{pjbr} is only correct for two-qubit states, where $T(\rho)$ is a matrix with entries given by $T_{i,j}(\rho)=Tr[\rho(\sigma_i\otimes \sigma_j)]$, we have presented a new approach in derivations. Our
innovation also lies in choosing the POVM measurement operators and
in calculating the expected CHSH values. Our approach can be also
extended to the case of mixed states.

We have also investigated the shareability of genuine
tripartite nonlocality. For the generalized GHZ state, it
has been shown that from our measurement schemes two Charlies can
share the genuine nonlocality with Alice and Bob. Our results may
also highlight researches on sharing general multipartite quantum
nonlocalities and other quantum correlations such as quantum
steerability \cite{sdsa,sgag}, entanglement \cite{asau,adaa} and
coherence \cite{sasm}.

Finally, in this article we have constructed the higher-dimensional dichotomic POVM measurement operators in terms of the Pauli operators. As the Pauli operators are easily implemented in experiments, the POVM operators we constructed may have potential advantages in some specific experimental implementations \cite{mzxc,mlmg}. Our approach may also highlight the related applications in randomness generation \cite{fmrm}, quantum teleportation \cite{sasa} and random access codes  \cite{kanb}. It would be also interesting to give some insight into larger
multipartite scenarios and explore the relationship between our methodology and, for instance, nonlocality depth sharing  \cite{fjan} and monogamy relations \cite{rmmj}.

\bigskip
Acknowledgments:  This work is supported by the National Natural
Science Foundation of China under Grant Nos. 11861031, 11675113 and
12075159, Beijing Municipal Commission of Education under grant No.
KZ201810028042, Beijing Natural Science Foundation (Z190005), the
Education Department of Hainan Province, project number
Hnky2020ZD-10, Academy for Multidisciplinary Studies, Capital Normal
University and Shenzhen Institute for Quantum Science and
Engineering, Southern University of Science and Technology (Grant
No. SIQSE202005).

\bigskip
\section*{APPENDIX}
\setcounter{equation}{0} \renewcommand%
\theequation{A\arabic{equation}}

\subsection{Proof of Theorem 1}

We first consider the simplest case that both $t$ and $s$ are even.
From (\ref{a00o}), (\ref{a01o}), (\ref{b00o}) and (\ref{b01o}) we
have
$$A_0=A_{0|0}-A_{1|0}=I_{\frac{s}{2}}\otimes(\cos\theta\sigma_3+\sin\theta\sigma_1),$$
$$A_1=A_{0|1}-A_{1|1}=I_{\frac{s}{2}}\otimes(\cos\theta\sigma_3-\sin\theta\sigma_1)$$
and
$$B_0^{(k)}=B_{0|0}-B_{1|0}=I_{\frac{t}{2}}\otimes\sigma_3,$$
$$B_1^{(k)}=B_{0|1}-B_{1|1}=I_{\frac{t}{2}}\otimes\gamma_k\sigma_1,$$
we have
\begin{eqnarray}\label{s4} & &I_{CHSH}^{(k)} \nonumber\\ & = & Tr[\rho_{AB}^{(k)}((A_0+A_1)\otimes B_0^{(k)})]\nonumber\\
&&+Tr[\rho_{AB}^{(k)}((A_0-A_1)\otimes B_1^{(k)})] \nonumber\\&=&2\cos\theta
 \ Tr[\rho_{AB}^{(k)}(I_{\frac{s}{2}}\otimes\sigma_3)\otimes(I_{\frac{t}{2}}\otimes\sigma_3)]\nonumber\\
& &+2\gamma_k\sin\theta \
Tr[\rho_{AB}^{(k)}(I_{\frac{s}{2}}\otimes\sigma_1)\otimes(I_{\frac{t}{2}}\otimes\sigma_1)].
\end{eqnarray}

For the case $k=1$, we get
\begin{eqnarray} \label{s5}& &I_{CHSH}^{(1)} \nonumber\\&=&2\cos\theta
\ Tr[\rho_{AB}^{(1)}(I_{\frac{s}{2}}\otimes\sigma_3)\otimes(I_{\frac{t}{2}}\otimes\sigma_3)]\nonumber\\
& &+2\gamma_1\sin\theta
\  Tr[\rho_{AB}^{(1)}(I_{\frac{s}{2}}\otimes\sigma_1)\otimes(I_{\frac{t}{2}}\otimes\sigma_1)]\nonumber\\&=&2\cos\theta(\Sigma_ic_i^2)+2\gamma_1(2(c_1c_2+c_3c_4+\cdots))\sin\theta\nonumber\\
&=&2\cos\theta+2\gamma_1L\sin\theta,
\end{eqnarray}
where we have used $\rho_{AB}^{(1)}=\Sigma_{i,j}c_ic_j|ii\rangle\langle jj|$ and $L=2(c_1c_2+c_3c_4+\cdots)$.

To obtain the relation between $\rho_{AB}^{(1)}$
and$\rho_{AB}^{(k)}$ we use the following equation,
\begin{eqnarray}&&\sqrt{\frac{1}{2}(I_t\pm\gamma_k I_{\frac{m}{2}}\otimes\sigma)}=\nonumber\\
&&\frac{(\sqrt{1+\gamma_k}+\sqrt{1-\gamma_k})I_t\pm(\sqrt{1+\gamma_k}-\sqrt{1-\gamma_k})
(I_{\frac{m}{2}}\otimes\sigma)}{2\sqrt{2}}\nonumber
\end{eqnarray}
for $\sigma=\sigma_1,\sigma_3$. By using the L\"uders update rule we have
\begin{eqnarray}\label{s6} & &\rho_{AB}^{(k)}=\frac{1}{2}\Sigma_{b,y}(I_s\otimes\sqrt{B_{b|y}^{(k-1)}})\rho_{AB}^{(k-1)}(I_s\otimes\sqrt{B_{b|y}^{(k-1)}})\nonumber\\
&=&\frac{1}{4}(I_s\otimes(I_\frac{t}{2}\otimes\sigma_3))\rho_{AB}^{(k-1)}(I_s\otimes(I_\frac{t}{2}\otimes\sigma_3))\nonumber\\
&+&\frac{1-\sqrt{1-\gamma_{k-1}^2}}{4}(I_s\otimes(I_\frac{t}{2}\otimes\sigma_1))\rho_{AB}^{(k-1)}(I_s\otimes(I_\frac{t}{2}\otimes\sigma_1))\nonumber\\
&+&\frac{2+\sqrt{1-\gamma_{k-1}^2}}{4}\rho_{AB}^{(k-1)}.
\end{eqnarray}
Substituting (\ref{s6}) into (\ref{s4}) and taking into account the relations
$\sigma_3\sigma_3\sigma_3=\sigma_3$ and
$\sigma_1\sigma_3\sigma_1=-\sigma_3$, we get
\begin{eqnarray}& & Tr[\rho_{AB}^{(k)}(I_{\frac{s}{2}}\otimes\sigma_3)\otimes(I_{\frac{t}{2}}\otimes\sigma_3)]\nonumber\\
&=&(\frac{1}{4}-\frac{1-\sqrt{1-\gamma_{k-1}^2}}{4}+\frac{2+\sqrt{1-\gamma_{k-1}^2}}{4})\nonumber\\
&\cdot&Tr[\rho_{AB}^{(k-1)}(I_{\frac{s}{2}}\otimes\sigma_3)\otimes(I_{\frac{t}{2}}\otimes\sigma_3)]\nonumber\\
&=&\frac{1+\sqrt{1-\gamma_{k-1}^2}}{2}
\nonumber\\
&\cdot&Tr[\rho_{AB}^{(k-1)}(I_{\frac{s}{2}}\otimes\sigma_3)\otimes(I_{\frac{t}{2}}\otimes\sigma_3)].\nonumber
\end{eqnarray}

Similarly we can obtain
\begin{eqnarray}& & Tr[\rho_{AB}^{(k)}(I_{\frac{s}{2}}\otimes\sigma_1)\otimes(I_{\frac{t}{2}}\otimes\sigma_1)]\nonumber\\
&=&\frac{1}{2}Tr[\rho_{AB}^{(k-1)}(I_{\frac{s}{2}}\otimes\sigma_1)\otimes(I_{\frac{t}{2}}\otimes\sigma_1)].\nonumber
\end{eqnarray}
By recursion, we get
\begin{eqnarray}\label{s7}
&&Tr[\rho_{AB}^{(k)}(I_{\frac{s}{2}}\otimes\sigma_3)\otimes(I_{\frac{t}{2}}\otimes\sigma_3)]=\nonumber\\
&&2^{1-k}\prod_{j=1}^{k-1}(1+\sqrt{1-\gamma_{j}^2})Tr[\rho_{AB}^{(1)}(I_{\frac{s}{2}}\otimes\sigma_3)\nonumber\\
&&\otimes(I_{\frac{t}{2}}\otimes\sigma_3)]
\end{eqnarray}
and
\begin{eqnarray}\label{s8}& & Tr[\rho_{AB}^{(k)}(I_{\frac{s}{2}}\otimes\sigma_1)\otimes(I_{\frac{t}{2}}\otimes\sigma_1)]=\nonumber\\
&&2^{1-k}Tr[\rho_{AB}^{(1)}(I_{\frac{s}{2}}\otimes\sigma_1)\otimes(I_{\frac{t}{2}}\otimes\sigma_1)].
\end{eqnarray}
Substituting (\ref{s7}) and (\ref{s8}) into (\ref{s4}) and using (\ref{s5}), we
obtain (\ref{s3}).

Now we prove the most complex case that both $s$ and $t$ are odd.
From (\ref{a00e}), (\ref{a01e}), (\ref{b00e}) and (\ref{b01e}) we
have
$$A_0=A_{0|0}-A_{1|0}=\left(\begin{array}{cc}
    I_{[\frac{s}{2}]}\otimes(\cos\theta\sigma_3+\sin\theta\sigma_1) &  0 \\
    0 & 1
  \end{array}\right),$$
$$A_1=A_{0|1}-A_{1|1}=\left(\begin{array}{cc}
    I_{[\frac{s}{2}]}\otimes(\cos\theta\sigma_3-\sin\theta\sigma_1) &  0 \\
    0 & 1
  \end{array}\right),$$
$$B_0^{(k)}=B_{0|0}-B_{1|0}=\left(\begin{array}{cc}
    I_{[\frac{t}{2}]}\otimes\sigma_3 &  0 \\
    0 & 1
  \end{array}\right)$$
and
$$B_1^{(k)}=B_{0|1}-B_{1|1}=\gamma_k\left(\begin{array}{cc}
    I_{[\frac{t}{2}]}\otimes\sigma_1 &  0 \\
    0 & 1
  \end{array}\right).
$$
Correspondingly, we have
\begin{eqnarray} & &I_{CHSH}^{(k)} \nonumber\\ & = & Tr[\rho_{AB}^{(k)}((A_0+A_1)\otimes B_0^{(k)})]\nonumber\\
&&+Tr[\rho^{(k)}((A_0-A_1)\otimes B_1^{(k)})] \nonumber\\
&=&2
  Tr\left[\rho_{AB}^{(k)}\left(\begin{array}{cc}
    I_{[\frac{s}{2}]}\otimes(\cos\theta\sigma_3) &  0 \\
    0 & 1 \\
  \end{array}\right)\otimes\left(\begin{array}{cc}
    I_{[\frac{t}{2}]}\otimes\sigma_3 &  0 \\
    0 & 1 \\
  \end{array}\right)\right]\nonumber\\
& &+2\gamma_k\sin\theta Tr\left[\rho_{AB}^{(k)}\left(\begin{array}{cc}
    I_{[\frac{s}{2}]}\otimes\sigma_1 &  0 \\
    0 & 0 \\
  \end{array}\right)\otimes\left(\begin{array}{cc}
    I_{[\frac{t}{2}]}\otimes\sigma_1 &  0 \\
    0 & 1 \\
  \end{array}\right)\right]\nonumber.
\end{eqnarray}
For the case $k=1$, one has
\begin{eqnarray} & &I_{CHSH}^{(1)} \nonumber\\&=&2
  Tr[\rho_{AB}^{(1)}\left(\begin{array}{cc}
    I_{[\frac{s}{2}]}\otimes(\cos\theta\sigma_3) &  0 \\
    0 & 1 \\
  \end{array}\right)\otimes\left(\begin{array}{cc}
    I_{[\frac{t}{2}]}\otimes\sigma_3 &  0 \\
    0 & 1 \\
  \end{array}\right)]\nonumber\\
& &+2\gamma_k\sin\theta Tr[\rho_{AB}^{(1)}\left(\begin{array}{cc}
    I_{[\frac{s}{2}]}\otimes\sigma_1 &  0 \\
    0 & 0 \\
  \end{array}\right)\otimes\left(\begin{array}{cc}
    I_{[\frac{t}{2}]}\otimes\sigma_1 &  0 \\
    0 & 1 \\
  \end{array}\right)]\nonumber\\
&=&\left\{
             \begin{array}{lr}
          2\cos\theta(1-c_s^2)+2c_s^2+2\gamma_1L\sin\theta, \ \ \mbox{if}\  s=t\  \mbox{and}\  c_s\neq 0,  \\[1mm]

          2\cos\theta+2\gamma_1L\sin\theta, \ \ \mbox{otherwise}
             \end{array}
\right. \nonumber\\
 &\geq & 2\cos\theta+2\gamma_1L\sin\theta,\nonumber
\end{eqnarray}
where we have used
$\rho_{AB}^{(1)}=\Sigma_{i,j=1}^sc_ic_j|ii\rangle\langle jj|$ and
$L=2(c_1c_2+c_3c_4+\cdots+c_{2[\frac{s}{2}]-1}c_{2[\frac{s}{2}]})$.

Using the following identity
\begin{eqnarray}& &\sqrt{\frac{1}{2}(I_t+\gamma_k\left(\begin{array}{cc}
    I_{[\frac{t}{2}]}\otimes\sigma &  0 \\
    0 & 1 \\
  \end{array}\right))}\nonumber\\
&=&\frac{(\sqrt{1+\gamma_k}+\sqrt{1-\gamma_k})I_t}{2\sqrt{2}}\nonumber\\
&\pm
&\frac{(\sqrt{1+\gamma_k}-\sqrt{1-\gamma_k})\left(\begin{array}{cc}
    I_{[\frac{t}{2}]}\otimes\sigma &  0 \\
    0 & 1 \\
  \end{array}\right)}{2\sqrt{2}}\nonumber
\end{eqnarray}
and repeating the similar
process for even $t$ and $s$, we get the relations similar to
Eq. (\ref{s7}) and Eq.(\ref{s8}). Moreover, we get
$$
I^{(k)}_{CHSH}\geq 2^{2-k}(\gamma_k
L\sin\theta+\cos\theta\prod_{j=1}^{k-1}(1+\sqrt{1-\gamma_j^2})).
$$

It is straightforward to prove that the above inequality holds also
for the cases of even (odd) $t$ and odd (even) $s$.

\subsection{Proof of Theorem 2}

The Theorem can be proved by using two lemmas in \cite{pjbr}. Since
the sequence $\gamma_k(\theta)\}_{k\in \mathbb{N}}$ defined by
(\ref{s11}) reduces to the ones given in \cite{pjbr} when
$L=\lambda_1$, we have that
$\{\gamma_1(\theta)=(1+\epsilon)\frac{1-\cos(\theta)}{L\sin(\theta)}$
is positive and an increasing function of $\theta$ for
$\theta\in(0,\frac{\pi}{4}]$ and $\varepsilon >0$. Moreover, the
subsequence $\{\gamma_k(\theta)\}_{k\in \mathbb{N}}$ consisting of
all finite terms is a strictly increasing sequence. In addition to
the sequence being monotonically increasing, each term in the
sequence also has a vanishing limit as $\theta$ approaches $0$. For
any $n\in \mathbb{N}$ there exists some
$\theta_n\in(0,\frac{\pi}{4}]$ such that for all $k\leq n$ and
$\theta\in(0,\theta_n)$, $\gamma_k(\theta) < 1$. Moreover, we have
$\lim_{\theta\rightarrow 0^+}\gamma_n(\theta)=0$ for all
$n\in\mathbb{N}$.

Therefore, there exists some $\theta_n\in(0,\frac{\pi}{4}]$
such that $\gamma_k(\theta) < 1$ and
$0<\gamma_1(\theta_n)<\gamma_2(\theta_n)<\cdots<\gamma_n(\theta_n)$,
with each $\gamma_k(\theta_n)$ satisfying the condition (\ref{s10})
and thus giving rise to Bell violations.

\subsection{Proof of Theorem 3}

First of all, we have
$$
\ba{l}
A_0=A_{0|0}-A_{1|0}=\sigma_1,\\
A_1=A_{0|1}-A_{1|1}=\sigma_2,
\ea
$$
$$
\ba{l}
B_0=B_{0|0}-B_{1|0}=\cos\theta\sigma_1-\sin\theta\sigma_2,\\
B_1=B_{0|1}-B_{1|1}=\cos\theta\sigma_1+\sin\theta\sigma_2
\ea
$$
and
$$
\ba{l}
C_0^{(k)}=C_{0|0}-C_{1|0}=\sigma_1,\\
C_1^{(k)}=C_{0|1}-C_{1|1}=\gamma_k\sigma_2.
\ea
$$
Then
\begin{eqnarray}\label{s16} & &S^{(k)} =  Tr[\mathbb{S} \rho_{ABC}^{(k)}]\nonumber\\ & =
&2\cos\theta
Tr[\rho_{ABC}^{(k)}\sigma_1\otimes\sigma_1\otimes\sigma_1]\nonumber\\
&-&2\sin\theta
Tr[\rho_{ABC}^{(k)}\sigma_1\otimes\sigma_2\otimes\sigma_2]
\nonumber\\&-&2\sin\theta\gamma_k
Tr[\rho_{ABC}^{(k)}\sigma_2\otimes\sigma_2\otimes\sigma_1]\nonumber\\&-&2\cos\theta\gamma_k
Tr[\rho_{ABC}^{(k)}\sigma_2\otimes\sigma_1\otimes\sigma_2].
\end{eqnarray}

For the case $k=1$, substituting $\rho_{ABC}^{1}=|\psi_g\rangle\langle\psi_g|$ into (\ref{s16})
we get
\begin{eqnarray}S^{(1)}&=&2\cos\theta\sin2\alpha+2\sin\theta\sin2\alpha\nonumber\\
&&+2\sin\theta\gamma_1\sin2\alpha+2\cos\theta\gamma_1\sin2\alpha\nonumber\\
&=&2\sin2\alpha(\cos\theta+\sin\theta)(1+\gamma_1).\nonumber
\end{eqnarray}
By using the following identity
\begin{eqnarray}& &\sqrt{\frac{1}{2}(I\otimes I+\gamma_k\sigma )}\nonumber\\
&=&\frac{(\sqrt{1+\gamma_k}+\sqrt{1-\gamma_k})I\otimes
I\pm(\sqrt{1+\gamma_k}-\sqrt{1-\gamma_k})\sigma}{2\sqrt{2}}\nonumber
\end{eqnarray}
for $\sigma=\sigma_1,\sigma_2$ and repeating the same
processes in the proof of Theorem 1, we obtain
\begin{eqnarray} &
&Tr[\rho_{ABC}^{(k)}\sigma_1\otimes\sigma_1\otimes\sigma_1]\nonumber\\
&=&(\frac{1}{4}-\frac{1-\sqrt{1-\gamma^2_{k-1}}}{4}+\frac{2+\sqrt{1-\gamma^2_{k-1}}}{4})
\nonumber\\&\cdot&
Tr[\rho_{ABC}^{(k-1)}\sigma_1\otimes\sigma_1\otimes\sigma_1]\nonumber\\
&=&(\frac{1+\sqrt{1+\gamma^2_{k-1}}}{2})Tr[\rho_{ABC}^{(k-1)}\sigma_1\otimes\sigma_1\otimes\sigma_1].\nonumber
\end{eqnarray}
Then
\begin{eqnarray} &
&Tr[\rho_{ABC}^{(k)}\sigma_1\otimes\sigma_1\otimes\sigma_1]=\nonumber\\
&&2^{1-k}Tr[\rho_{ABC}^{(1)}\sigma_1\otimes\sigma_1\otimes\sigma_1]\Pi_{j=1}^{k-1}(1+\sqrt{1-\gamma_j^2}).\nonumber
\end{eqnarray}
At last we have
\be
S^{(k)}=2^{2-k}\sin2\alpha(\cos\theta+\sin\theta)(\gamma_k
+\prod_{j=1}^{k-1}(1+\sqrt{1-\gamma_j^2})).\nonumber
\ee

\subsection{Proof of Theorem 4}

From (\ref{s13}) Alice, Bob and Charlie share a genuine nonlocally correlated state if
\begin{eqnarray}\label{s14}& &\gamma_k>\frac{2^{k}}{\sin2\alpha(\cos\theta+\sin\theta)}-\prod_{j=1}^{k-1}(1+\sqrt{1-\gamma_j^2}).\nonumber
\end{eqnarray}
Let $\varepsilon > 0$ and
$\gamma_1(\theta)=(1+\epsilon)(\frac{2}{\sin2\alpha(\cos\theta+\sin\theta)}-1)$.
For $k\in \{2,\cdots,n\}$ recursively set
$$\gamma_k(\theta)=\left\{
             \begin{array}{lr}
          (1+\epsilon)(\frac{2^{k}}{\sin2\alpha(\cos\theta+\sin\theta)}-P_k),~ \mbox{if}~ \gamma_{k-1}(\theta)\in(0,1),  \\
          \infty, \ \ \mbox{otherwise},
          \end{array}\right.
$$
where $P_k=\prod_{j=1}^{k-1}(1+\sqrt{1-\gamma_j(\theta)^2})$.

Suppose there is a finite integer number $m\geq 2$, such that $0<\gamma_j(\theta)<1$
for all $j=1,2,\cdots,m$. Then
$1<1+\sqrt{1-\gamma_j(\theta)^2}<2$. The bound
$1+\sqrt{1-\gamma_j^2(\theta)}<2$ implies
that $\gamma_j(\theta)/\gamma_{j-1}(\theta)>2$.

Take $\sin^22\alpha=\frac{8}{9}$. We have
$$
\gamma_1(\theta)>\gamma_1=\frac{3\sqrt{2}}{4}\frac{2}{\cos\theta+\sin\theta}-1.
$$
Because $1\leq\cos\theta+\sin\theta\leq\sqrt{2}$,
$\gamma_1(\theta)\in [\frac{1}{2},1]$ for some
$\theta\in[0,\frac{\pi}{4}]$. In this case,
$\gamma_2(\theta)>2\gamma_1(\theta) > 1$. Thus, there does not exist
$\theta$ such that $\gamma_2(\theta)<1$. As $\gamma_1$ is a
monotonically decreasing function of $\sin^22\alpha$, when
$\sin^22\alpha > \frac{8}{9}$, $\gamma_1(\theta)$ could be less than
$\frac{1}{2}$. The smallest value of $\gamma_1(\theta)$ is
$\sqrt{2}-1$. In this case there exits some $\theta$ such that
$1>\gamma_2(\theta)>2\gamma_1(\theta)$, which completes the proof of
Theorem 4.
\end{document}